\documentstyle[12pt,amssymb]{article}
\newcommand{\be}{\begin{equation}}
\newcommand{\ee}{\end{equation}}
\newcommand{\bear}{\begin{eqnarray}}
\newcommand{\ear}{\end{eqnarray}}
\date{}
\renewcommand{\theequation}{\arabic{section}.\arabic{equation}}
\newcommand{\eins}{  1\!{\rm l}  }

\newcommand{\cL}{{\cal L}}
\newcommand{\RRe}{\rm{Re}}
\newcommand{\RIm}{\rm{Im}}
\def\vec#1{\mathchoice{\mbox{\boldmath$\mathrm\displaystyle#1$}}
{\mbox{\boldmath$\mathrm\textstyle#1$}}
{\mbox{\boldmath$\mathrm\scriptstyle#1$}}
{\mbox{\boldmath$\mathrm\scriptscriptstyle#1$}}}
\newcommand{\bm}[1]{\mbox{\boldmath$#1$}}  
\renewcommand{\vec}{\bm}

\begin{document}
\begin{flushright}
HD-THEP-98-56\\
PITHA 98/40
\end{flushright}
\vspace{1.5cm}

\begin{center}
{\LARGE
Flavour Dynamics with General Scalar Fields\footnote{supported by BMBF,
contracts 05 7AC 9EP and 05 6HD 91P(0)}
}\\
\vspace{1cm}
by\\
\vspace{.5cm}
{\sc W. Bernreuther\footnote{breuther@physik.rwth-aachen.de}
}\\
\medskip
{\em Institut f{\"u}r Theoretische Physik} \\
{\em RWTH Aachen, D-52056 Aachen, Germany}\\
\vspace{.5cm}
and \\
\vspace{.5cm}
{\sc O. Nachtmann
\footnote{O.Nachtmann@thphys.uni-heidelberg.de}
}
\\
\medskip
{\em Institut f{\"u}r Theoretische Physik} \\
{\em Universit\"at Heidelberg, D-69120 Heidelberg, Germany}

\end{center}
\vspace{1cm}

\begin{abstract}
We consider a spontaneously broken gauge theory based on the
standard model (SM) group $G = SU(2)\times U(1)$ with scalar
fields that carry arbitrary representations of $G$, and we investigate
some general properties of the charged and neutral current involving these
fields.  In particular we derive the conditions for having
real or complex couplings of the $Z$ boson to two different
neutral or charged scalar fields, and for the existence of CP-violating
$Z$-scalar-scalar couplings. Moreover, we study models with the same
fermion content as in the SM, with one $SU(2)$ Higgs singlet, and an arbitrary
number of Higgs doublets. We show that the structure of the $Z$-Higgs boson
and of the Yukawa couplings in these models can be such that
CP-violating $Zb{\bar b}G$ form factors which conserve chirality
are induced at the one-loop level.
\end{abstract}

\newpage

\section{Introduction}
The standard model (SM) of elementary particle physics \cite{1}
has been very successful, so far, when compared to experiments.
For instance LEP1 and SLC, with its precision data,  have proved to be
an ideal testing ground of the SM, where the theory, including
its quantum corrections, has been checked (for recent reviews, see
\cite{2,3}). However,
one crucial aspect of the SM has remained practically unexplored
experimentally till to date: the electroweak symmetry breaking sector.
In the standard picture an elementary scalar field\footnote{Other scenarios for
electroweak symmetry breaking like technicolor models have
been discussed \cite{5}, but remain less well-developed theoretically
and seem to be disfavoured by the data; see, e.g.,  \cite{Ellis}}
is responsible
for spontaneous breaking of the electroweak gauge group
$G=SU(2)\times U(1)$ and for the generation of particle masses \cite{4,7}.
 However, extensions of the SM, for which there are a
number of well-known theoretical motivations, almost invariably
entail a larger scalar field content than in the SM. 
That is,  additional Higgs fields,
but possibly also scalar leptoquarks or, in supersymmetric extensions
of the SM, squarks and sleptons.

In this article we shall investigate an $SU(2)\times U(1)$ gauge
theory
with an arbitrary number of scalar fields.
For ease of notation we will collectively
denote these fields as Higgs fields.
Our aim is to answer some general
questions concerning the charged and in particular the
neutral current involving these fields, namely:

What are the conditions for having  a real or complex coupling of the
$Z$ boson to two different neutral or charged physical
Higgs fields?

Can there be CP-violating $Z$-Higgs couplings? What are the conditions
that complex phases in such couplings can or cannot be ``rotated away''?

Our article is organised as follows. In section 2 we introduce the
general Higgs field and discuss spontaneous symmetry breaking.
In section 3 we study the question of non-diagonal $Z$-Higgs-Higgs boson
couplings. In section 4 we apply the general formalism to
models with fermion content as in the SM and  with one $SU(2)$
Higgs singlet and any
number $l$ of Higgs doublets.
In Section 5 we show how  such models with $l\geq 3$
Higgs doublets  provide all the prerequisites
for generating
chirality-conserving, CP-violating effective $Zb\bar bG$ couplings
at the one-loop level.
Section 6 contains our conclusions. In the appendices we discuss some
properties of the general $SU(2)\times U(1)$ representation
carried by scalar fields.

\section{The general Higgs field and spontaneous symmetry breaking}

We consider a gauge theory based on the electroweak gauge
group
$G=SU(2)\times U(1)$ (for our notation cf. \cite{6}).
The elements of $G$ will be denoted by $U$.
A suitable concrete realization of $G$ is by $2\times2$ matrices with
the following parametrization:
\be\label{2.1}
U(\vec{\varphi},\psi)=\exp\left[i\frac{1}{2}\tau_a\varphi_a+
iy_0\psi\right]\ee
where $\tau_a\ (a=1,2,3)$ are the Pauli matrices and $\vec\varphi=
(\varphi_a)$ is restricted to
\be\label{2.2}
|\vec\varphi|<2\pi.\ee
We assume $y^{-1}_0$ to be a natural number
$(y_0^{-1}=1,2,...)$ which will be chosen conveniently
later on, and we have
\be\label{2.3}
|\psi|<\pi y_0^{-1}.\ee
The parametrization (\ref{2.1}) is almost everywhere regular
on $G$. For our purposes below it suffices to note that
(\ref{2.1}) is regular in a suitable neighbourhood of the
unit element of $G$: $U={\eins}_2$.
Taking an arbitrary element $U_0\in G$, we get a parametrization
of the elements of $G$ which is regular in a neighbourhood
of $U_0$ by setting
\be\label{2.4}
U=U(\vec\varphi,\psi)\cdot U_0.\ee
{}From (\ref{2.1}) to (\ref{2.4})
we see that $G$ is a differentiable, compact manifold.

In the following $\chi$ denotes a Higgs field that transforms
under $G$ according to an arbitrary representation, which is
in general reducible and contains real orthogonal as well as
complex unitary parts. Let us first show that without loss of generality
we can assume $\chi$ to carry a real orthogonal representation
of $G$. To see this, consider a Higgs field\footnote{Here and in the
following we suppress the space-time variable $x$, if there
is no danger of misunderstanding. Thus  $\phi\in {\mathbb{C}}_r$
in (\ref{2.5}) and below is to be read as $\phi(x)\in {\mathbb{C}}_r$
for each $x$. Likewise we introduce in (\ref{2.8}) a $2r$ component
real vector $\chi(x)$.} 
$\phi$ carrying a unitary
representation
of dimension $r$:
\be\label{2.5}
\phi=\left(\begin{array}{c} \phi_1\\ .\\ .\\ .\\ \phi_r
\end{array}\right),\quad \phi\in {\mathbb{C}}_r\ee
where the action of $G$ is as follows:
\be\label{2.6}
U:\quad \phi\longrightarrow D_r(U)\phi,\ee
\bear\label{2.7}
&&D^\dagger_r(U) D_r(U)={\eins}_r,\nonumber\\
&&U\in G.\ear
We define a corresponding 
$2r$ component real Higgs field $\chi$  by setting
\bear\label{2.8}
&&\chi=(\chi_{\alpha, j})\nonumber\\
&&(\alpha=1,2;\quad j=1,.., r),\nonumber\\
&&\chi_{1,j}:=\RRe\phi_j,\nonumber\\
&&\chi_{2,j}:=\RIm\phi_j.\ear
Furthermore we define the real $2r\times 2r$ matrices
\be\label{2.9}
R_{2r}(U):={\eins}_2\otimes {\RRe} D_r(U)-\epsilon\otimes {\RIm}
\ D_r(U)\ee
where
\be\label{2.10}
\epsilon=\left(\begin{array}{cc}
0& 1\\
-1& 0\end{array}\right).\ee
It is easy to see that we have
\be\label{2.11}
\phi^\dagger \phi=\chi^T\chi.\ee
The transformation (\ref{2.6}) corresponds to
\be\label{2.12}
U:\ \chi\longrightarrow R_{2r}(U)\chi\ee
and
\[U\longrightarrow R_{2r}(U)\]
is a real orthogonal representation of $G$:
\bear\label{2.13}
&& R_{2r}(U)R_{2r}(U')=R_{2r}(U U'),\nonumber\\
&& R_{2r}(U^\dagger)=R_{2r}^T(U),\nonumber\\
&&R_{2r}^T(U)R_{2r}(U)={\eins}_{2r},\nonumber\\
&&(U,U'\in G).\ear

We can thus start with a general real Higgs field $\chi$ with
$n$ components
\be\label{2.14}
\chi=\left(\begin{array}{c}
\chi_1\\
.\\.\\.\\ \chi_n\end{array}\right),\ee
carrying an orthogonal representation of $G$:
\bear\label{2.15}
U:&&\chi\longrightarrow R(U)\chi,\nonumber\\
&&R^T(U)R(U)={\eins}.\ear
For infinitesimal transformations $U(\delta\vec\varphi,\delta\psi)$
(cf. (\ref{2.1})) we have
\be\label{2.16}
R(U(\delta\vec\varphi,\delta\psi))={\eins}+\delta\varphi_a
\tilde T_a+\delta\psi\tilde Y.\ee
Here $\tilde T_a$ and $\tilde Y$ are the real antisymmetric
matrices representing the generators of $G$:
\bear\label{2.18}
&&\tilde T_a+\tilde T_a^T=0,\nonumber\\
&&\tilde Y+\tilde Y^T=0;\ear
\bear\label{2.19}
&&[\tilde T_a,\tilde T_b]=-\epsilon_{abc}\tilde T_c,\nonumber\\
&&[\tilde Y,\tilde T_a]=0.\ear
We define the corresponding generator of the electromagnetic
gauge group $U(1)_{em}$ as usual by
\be\label{2.20}
\tilde Q=\tilde T_3+\tilde Y.\ee
We note the commutation relations following from (\ref{2.19})
and (\ref{2.20}):
\bear\label{2.21}
&&[\tilde Q,\tilde T_1]=-\tilde T_2,\nonumber\\
&&[\tilde Q,\tilde T_2]=\tilde T_1,\nonumber\\
&&[\tilde Q,\tilde T_3]=0.\ear

Consider next the matrix $\tilde Y^T\tilde Y$ which is symmetric and
positive semi-definite. We assume that the eigenvalues $y_j^2$ of
$\tilde Y^T\tilde Y$ satisfy
\bear\label{2.210}
|y_j|.y^{-1}_0\ &=&\ {\rm integer}\nonumber\\
(j&=&1,...,n).\ear
The representation of $G$  in the space of Higgs fields
is then single-valued (cf. Appendix A).

The reason for going through these subtleties here is because we
want to avoid the case where the Higgs representation of $G$ is only
single-valued when considered as a representation of the
universal covering group of $G$, which is not a compact manifold.
This would happen if the ratio of two numbers $|y_j|$ and $|y_k|$
was irrational.

We shall now study the Higgs part of the Lagrangian describing
an $SU(2)\times U(1)$ gauge theory with the arbitrary Higgs field
$\chi$:
\be\label{2.22}
{\cal L}_\chi=\frac{1}{2}(D_\mu\chi)^T(D^\mu\chi)-
V(\chi).\ee
Here
\be\label{2.23}
D_\mu\chi:=(\partial_\mu+gW^a_\mu\tilde T_a+g'B_\mu\tilde Y)\chi\ee
is the covariant derivative of $\chi$ and $g$ and $g'$ are the
$SU(2)$ and $U(1)$ coupling constants, respectively. The gauge boson fields
are denoted by $W_\mu^a$ and $B_\mu$.
The Higgs potential $V(\chi)$ must be invariant under $G$ and
is constrained by the requirements of hermiticity and renormalizabilty.
Thus $V$ can contain up to fourth powers of $\chi$.

In the following we let $V$ largely unspecified apart from
assuming that it leads to spontaneous symmetry
breaking where only the electromagnetic gauge group
$U(1)_{em}$ remains unbroken.
Let $v$ be the vector of vacuum expectation values of $\chi$ (at tree
level):
\be\label{2.24}
v=<0|\chi|0>\not=0.\ee
We must then have
\be\label{2.25}
\tilde Q v=0,\ee
and the three vectors
\be\label{2.26}
\tilde T_av\quad (a=1,2,3)\ee
must be linearly independent.
Using (\ref{2.18})-(\ref{2.21}) it is easy to derive the
following relations:
\bear
v^T\tilde T_av&=&0,\quad (a=1,2,3)\label{2.27},\\
v^T\tilde T_1\tilde T_1 v&=&v^T\tilde T_2\tilde T_2 v,\nonumber\\
v^T\tilde T_1\tilde T_2 v&=&-v^T\tilde T_2\tilde T_1v,\label{2.28}\\
v^T\tilde T_3\tilde T_3 v&=&v^T\tilde Y\tilde Yv,\nonumber\\
v^T\tilde T_3\tilde Y v&=&-v^T\tilde Y\tilde Y v.\label{2.29}\ear

Let us next define the shifted Higgs field $\chi'$ by
\be\label{2.30}
\chi':=\chi-v.\ee
We get then
\bear\label{2.31}
D_\mu\chi&=&(\partial_\mu+\tilde\Omega_\mu)\chi\nonumber\\
&=&\tilde\Omega_\mu v+\partial_\mu\chi'+\tilde\Omega_\mu\chi',\ear
where
\be\label{2.32}
\tilde\Omega_\mu:=gW^a_\mu\tilde T_a+g'B_\mu\tilde Y.\ee
In terms of the physical vector boson fields $Z_\mu,A_\mu$ we have
$(s\equiv \sin\theta_w,\ c\equiv\cos\theta_w,\ e=g s=g' c)$:
\bear\label{2.33}
&&Z_\mu=cW^3_\mu-sB_\mu,\nonumber\\
&&A_\mu=sW^3_\mu+cB_\mu;\ear
\be\label{2.34}
\tilde \Omega_\mu=\frac{e}{s}\left(W_\mu^1\tilde T_1+W_\mu^2
\tilde T_2\right)
+\frac{e}{sc}Z_\mu\left(\tilde T_3-s^2\tilde Q\right)
+e A_\mu\tilde Q.\ee
With this the Lagrangian ${\cal L}_\chi$ (\ref{2.22})
reads:
\bear\label{2.35}
{\cal L}_\chi&=&\frac{1}{2} v^T\tilde\Omega^T_\mu
\tilde\Omega^\mu v
+\frac{1}{2}\partial_\mu{\chi'}^T\partial^\mu\chi'
-v^T\tilde\Omega_\mu\partial^\mu\chi'\nonumber\\
&&-{\chi'}^T\tilde\Omega_\mu\partial^\mu\chi'
-v^T\tilde\Omega_\mu\tilde\Omega^\mu\chi'
+\frac{1}{2}{\chi'}^T\tilde\Omega^T_\mu\tilde\Omega^\mu\chi'
-V(v+\chi').\ear
The successive terms on the r.h.s. of (\ref{2.35}) will be
denoted by ${\cal L}_\chi^{(i)},\
i=1,...,7$.

Aspects of such a Lagrangian (\ref{2.35}) have
been studied previously, for instance the $Z-W$-Higgs coupling in
\cite{8,9},  Higgs triplets in \cite{10,11}, and radiative corrections
for models with Higgs triplets in \cite{12,Hollik}.

Let us first study the term bilinear in the vector boson and
Higgs fields in (\ref{2.35}).
\bear\label{2.36}
{\cal L}^{(3)}_\chi:&=&-v^T\tilde\Omega_\mu\partial^\mu\chi'\nonumber\\
&=&-\frac{e}{s}\left[W^1_\mu\partial^\mu(v^T\tilde T_1\chi')
+W^2_\mu\partial^\mu(v^T\tilde T_2\chi')\right]
-\frac{e}{sc}Z_\mu\partial^\mu(v^T\tilde T_3\chi').\ear
In order to discuss the particle content and the couplings of
physical particles, it is convenient to use the unitary gauge
(for a review and an extensive use of this gauge cf. \cite{7}),
which is defined by the condition:
\be\label{2.37}
v^T\tilde T_a\chi'=0\quad {\rm for}\quad a=1,2,3.\ee
Can (\ref{2.37}) always be met?
An affirmative answer to this question was given in \cite{7} for the case of  
a compact
group. The proof of \cite{7} goes through also in our case since $G$
is also  compact and, by the condition (\ref{2.210}), we have
excluded multivalued representations of $SU(2)\times U(1)$ which
would force us to go to the non-compact universal covering
group.

To recall the construction of \cite{7} we note first that due to
(\ref{2.27}) the condition (\ref{2.37}) is
equivalent to
\be\label{2.38}
v^T\tilde T_a\chi=0\quad{\rm for}\quad a=1,2,3.\ee
Since $v^T\tilde T_a$ $(a=1,2,3)$
are linearly independent (cf. (\ref{2.26})), (\ref{2.38}) defines
a $n-3$ dimensional linear subspace ${\mathbb{R}}_{n-3}\subset {\mathbb{R}}_n$.
\be\label{2.39}
{\mathbb{R}}_{n-3}=\{\chi\ |\ v^T\tilde T_a\chi=0\quad
{\rm for}\quad a=1,2,3\}.\ee
Let now $\chi$ be an arbitrary vector in ${\mathbb{R}}_n$ and consider the
following real function on the compact manifold $G$
\bear\label{2.40}
U\to f(U)&=&v^TR(U)\chi\nonumber\\
(U&\in&G).\ear
$U_0\in G$ which maximises $f(U)$
transforms the arbitrary vector $\chi\in{\mathbb{R}}_n$ into a
vector $R(U_0)\chi$ lying in the subspace ${\mathbb{R}}_{n-3}$ (\ref{2.39})
of the vectors compatible with the gauge condition. Thus the bilinear
couplings of $W^{1,2}_\mu(x),\ Z_\mu(x)$ to the Higgs field
$\chi(x)$ in (\ref{2.36}) can at any space time point $x$ be rotated
away by a suitable transformation $U_0(x)\in G$,
where $U_0(x)$ will in general, of course, depend on $x$.
In Appendix B we discuss further some properties of the gauge orbits
of our scalar fields.

Having disposed of the bilinear vector boson-Higgs field part
${\cal L}_\chi^{(3)}$ of ${\cal L}_\chi$ (\ref{2.35})
with the help of the gauge condition (\ref{2.37}) we turn next
to the term ${\cal L}^{(1)}_\chi$, bilinear in the vector boson
fields, i.e., the vector boson mass term. From (\ref{2.35}), (\ref{2.34})
and using (\ref{2.28}), (\ref{2.29}), we find
\bear\label{2.41}
{\cal L}_\chi^{(1)}&=&\frac{1}{2} v^T\tilde\Omega^T_\mu
\tilde\Omega^\mu v\nonumber\\
&&=\frac{1}{2}\left(\frac{e}{s}\right)^2
\frac{1}{2}v^T(\tilde T_1^T\tilde T_1
+\tilde T_2^T\tilde T_2)v
\cdot (W^1_\mu W^{1\mu}+W^2_\mu W^{2\mu})\nonumber\\
&&+\frac{1}{2}\left(\frac{e}{sc}\right)^2v^T\tilde T_3
^T\tilde T_3v\cdot Z_\mu Z^\mu.\ear
{}From (\ref{2.41}) we can read off the $W$ and
$Z$ masses (at tree level)
\bear\label{2.42}
m^2_W&=&\left(\frac{e}{s}\right)^2\frac{1}{2}v^T(\tilde T_1^T
\tilde T_1+\tilde T_2^T\tilde T_2)v\nonumber\\
&=&\left(\frac{e}{s}\right)^2\frac{1}{2}v^T(\tilde T_a^T
\tilde T_a-\tilde Y^T\tilde Y)v,\nonumber\\
m^2_Z&=&\left(\frac{e}{sc}\right)^2v^T\tilde T_3^T
\tilde T_3v\nonumber\\
&=&\left(\frac{e}{sc}\right)^2v^T\tilde Y^T
\tilde Yv.\ear
This result is of course well known.

Using the decomposition of the representation $U\to R(U)$ defined
in (\ref{2.15}) but considered as a unitary
representation in ${{\mathbb{C}}}_n$ as explained in appendix A, we can write
(\ref{2.42}) as follows (cf. (\ref{A.8})-(\ref{A.17})):
\bear\label{2.43}
m_W^2&=&\left(\frac{e}{s}\right)^2\frac{1}{2}\sum_{t,y}
[t(t+1)-y^2]v^T\ {\mathbb{P}}(t,y)v,\nonumber\\
m_Z^2&=&\left(\frac{e}{sc}\right)^2\sum_{t,y}y^2
v^T\ {\mathbb{P}}(t,y)v,\ear
where ${\mathbb{P}}(t,y)$ is the projector on the subspace with
representation $(t,y)$ of $G$. Here $t$ and $y$ are the
isospin and hypercharge quantum numbers, respectively.
As we see from (\ref{A.17}), only representations with
$y=-t,\ -t+1,..., t$ can contribute with nonzero weight
$v^T\ {\mathbb{P}}(t,y)v\not=0$ in the sums (\ref{2.43}). For the convenience
of the reader we have listed in Table 1
the values of $(t,y)$ for $t\leq3$
satisfying the above condition
and the corresponding values for the $\rho$ parameter, defined as
usual:
\be\label{2.44}
\rho:=\frac{m^2_W}{m^2_Z\cos^2\theta_W}.\ee
{}From (\ref{2.43}) we get
\be\label{2.45}
\rho=\frac{\sum_{t,y}[t(t+1)-y^2]v^T\ {\mathbb{P}}(t,y)v}
{2\sum_{t,y}y^2v^T\ {\mathbb{P}}(t,y)v}.\ee
Due to the non-negativity of the weights $v^T\ {\mathbb{P}}
(t,y)v\ge0$ (cf. (\ref{A.18}))
the value of $\rho$ for an arbitrary representation of $G$ must be
inside the interval spanned by the values of $\rho$
from the irreducible representations contributing with
nonzero weight in (\ref{2.43}).
Note  that the tree-level relation $\rho=1$
holds, apart from the  Higgs doublet representations
$t=1/2,\ y=\pm1/2$,
also for the triplet representations $t=3,\ y=\pm2$. (Actually,
there are other, higher-dimensional representations satisfying this
tree level relation; see for instance \cite{9}.)

\section{The general structure of the $Z$-Higgs-Higgs and
$W$-Higgs-Higgs vertices}
\setcounter{equation}{0}

In this section we derive some properties of the vertices describing
the coupling of the $Z$ and $W$ bosons to two physical Higgs particles.
In particular we give the conditions for having a non-diagonal
real or complex $Z$-Higgs-Higgs boson coupling.

The corresponding term of the Lagrangian (\ref{2.35})  is
\bear\label{3.1}
{\cal L}_\chi^{(4)}&=&-{\chi'}^T\tilde\Omega_\mu
\partial^\mu \chi'\nonumber\\
&=&-\frac{e}{s}W^1_\mu{\chi'}^T\tilde T_1\partial^\mu\chi'
-\frac{e}{s}W^2_\mu{\chi'}^T\tilde T_2\partial^\mu\chi'
\nonumber\\
&&-\frac{e}{sc}Z_\mu{\chi'}^T(\tilde T_3-s^2\tilde Q)\partial^\mu
\chi'-eA_\mu {\chi'}^T\tilde Q\partial^\mu\chi',\ear
where $\chi'$ is the shifted Higgs field (cf. (\ref{2.30})),
a vector  in the space ${\mathbb{R}}_{n-3}$ (\ref{2.39}). It is convenient
to introduce the projector onto this space of
the physical Higgs fields.
For this we define 3 vectors $w_a\ (a=1,2,3)$ in ${\mathbb{R}}_n$:
\bear\label{3.2}
w_a:&=&\tilde T_a v\cdot (v^T\tilde T_a^T\tilde T_av)^{-1/2}\nonumber\\
&&(\rm no\ summation\ over\ a).\ear
{}From (\ref{2.28}), (\ref{2.29}) and (\ref{2.42}) we find
\bear\label{3.3}
w_1&=&\tilde T_1 v\cdot\frac{e}{sm_W},\nonumber\\
w_2&=&\tilde T_2 v\cdot\frac{e}{sm_W},\nonumber\\
w_3&=&\tilde T_3 v\cdot\frac{e}{scm_Z},\ear
\be\label{3.4}
w^T_aw_b=\delta_{ab}.\ee
{}From (\ref{2.39}) we see that the vectors $w_a$ are the
normalised vectors orthogonal to ${\mathbb{R}}_{n-3}$.
The projector onto ${\mathbb{R}}_{n-3}$ is thus given by
\be\label{3.5}
{\mathbb{P}}':={\eins}-w_aw_a^T.\ee
{}From (\ref{2.21}) and (\ref{2.25}) we get
\bear\label{3.6}
&&\tilde Q w_1=-w_2,\nonumber\\
&&\tilde Q w_2=w_1,\nonumber\\
&&\tilde Q w_3=0,\ear
which leads to
\be\label{3.7}
[{\mathbb{P}}',\tilde Q]=0.\ee
Thus ${\mathbb{P}}'$ commutes with
the generator of electric charge. In general, however, ${\mathbb{P}}'$ will
\underbar{not} commute with $\tilde T_3$.

When discussing the couplings in (\ref{3.1})
we have to restrict the coupling
matrices $\tilde T_a,\tilde Q$ to the space of physical
Higgs fields. This can be done with the help of the
projector ${\mathbb{P}}'$ (\ref{3.5}). We define the following matrices
\bear\label{3.9}
&&\tilde T_a'':=\ {\mathbb{P}}'\tilde T_a{\mathbb{P}}'\quad (a=1,2,3),\nonumber\\
&&\tilde Q'':=\ {\mathbb{P}}'\tilde Q\ {\mathbb{P}}'.\ear
The matrices (\ref{3.9}) are block-diagonal, with
non-trivial $(n-3)\times(n-3)$ submatrices
$\tilde T_a',\tilde Q'$ on ${\mathbb{R}}_{n-3}$ and zero on
its orthogonal complement. In the following we shall only
deal with the submatrices on ${\mathbb{R}}_{n-3}$.
Eqs. (\ref{2.21}) and (\ref{3.7}) imply that
\be\label{3.10}
{}[\tilde T_3',\tilde Q']=0.\ee
Similarly we find
\bear\label{3.10a}
&&[\tilde Q',\ \tilde T_1']=-\tilde T_2',\nonumber\\
&&[\tilde Q',\ \tilde T_2']=\tilde T_1'.\ear
Note, however, that the matrices $\tilde T_a'$ $(a=1,2,3)$ will in
general \underbar{not} satisfy the $SU(2)$ commutation
relations.
Our aim is now to diagonalise the matrices $\tilde Q'$ and $\tilde T_3'$
and to arrange the components of the physical Higgs field
$\chi'$ into charge eigenstates. However, since
$\tilde Q'$ and $\tilde T_3'$ are antisymmetric real matrices
on the real space ${\mathbb{R}}_{n-3}$, this requires some nontrivial
work similar to the one of  Appendix A.

Let us embed the space ${\mathbb{R}}_{n-3}$ into the complex space
${\mathbb{C}}_{n-3}$
and define matrices:
\bear\label{3.11}
T_a'&=&\frac{1}{i}\tilde T_a',\quad (a=1,2,3),\nonumber\\
Q'&=&\frac{1}{i}\tilde Q'.\ear
We have
\bear\label{3.12}
\tilde T_a^{'T}&=&-\tilde T_a',\nonumber\\
{\tilde Q}^{'T}&=&-\tilde Q';\ear
\bear\label{3.13}
T_a'&=&-{T'}^T_a={T'}^\dagger_a=-T^{'*}_a,\nonumber\\
Q'&=&-{Q'}^T={Q'}^\dagger=-Q^{'*};\ear
\be\label{3.14}
[T_3',Q']=0.\ee
It follows from (\ref{3.13}), (\ref{3.14}) that the hermitian matrices
$Q', T_3'$ can be diagonalized simultaneously in
${\mathbb{C}}_{n-3}$. Let the eigenvalue pairs  be
$(q,t_3')$. We consider the double resolvent
\bear\label{3.15}
\frac{1}{(\xi-Q')(\eta-T_3')}
&=&\sum_{q,t_3'}\frac{{\mathbb{P}}(q,t_3')}{(\xi-q)(\eta-t_3')}\nonumber\\
&&(\xi,\eta\in{\mathbb{C}}),\ear
where ${\mathbb{P}}(q,t_3')$ is the projector onto the subspace of
eigenvectors associated with the eigenvalue pair $(q,t_3')$.
By taking the transposed of (\ref{3.15}) and using
(\ref{3.13}) we find
\be\label{3.16}
\frac{1}{(\xi+Q')(\eta+T_3')}=\sum_{q,t_3'}
\frac{{\mathbb{P}}^T(q,t_3')}{(\xi-q)(\eta-t_3')},\ee
\be\label{3.17}
\sum_{q,t_3'}\frac{{\mathbb{P}}(q,t_3')}{(\xi+q)(\eta+t_3')}
=\sum_{q,t_3'}\frac{{\mathbb{P}}^T(q,t_3')}{(\xi-q)(\eta-t_3')}.
\ee
Comparing the poles and residues on the r.h.s. and l.h.s.
of (\ref{3.17}), we conclude that with $(q,t_3')$ also
$(-q,-t_3')$ must be an eigenvalue pair and  $(q,t_3')$
and $(-q,-t_3')$ have the same multiplicity. For the
projectors we find
\be\label{3.18}
{\mathbb{P}}(-q,-t_3')={\mathbb{P}}^T(q,t_3').\ee

We treat now the eigenspaces with $q=0$ and with $q\not=0$
separately. Note that the eigenspace with $q=0$ must always
have dimension $\geq1$, since
\be\label{3.18a}
\chi'=c v\qquad (c\in {\mathbb{C}},\ c\not=0)\ee
is an eigenvector of $Q'$ in ${\mathbb{C}}_{n-3}$ with eigenvalue $0$:
\be\label{3.19}
Q'(c v)=0.\ee

Let us denote by $S_q$ the subspace of ${\mathbb{C}}_{n-3}$
corresponding to charge eigenvalue $q$ and arbitrary
$t_3'$. The projector onto $S_q$ is
\be\label{3.20}
{\mathbb{P}}(q)=\sum_{t_3'}{\mathbb{P}}(q,t_3').\ee
The hermiticity of ${\mathbb{P}}$ and
(\ref{3.18}) imply that ${\mathbb{P}}(0)$
 is a real, symmetric matrix
which can also be considered as a projector in the
real space ${\mathbb{R}}_{n-3}$. Hence real eigenvectors $u_1,...,
u_{r_0}\in {\mathbb{R}}_{n-3}$ exist, such that
\bear\label{3.22}
&&{\mathbb{P}}(0)u_j=u_j,\nonumber\\
&&Q'u_j=0,\nonumber\\
&&\qquad(j=1,...,r_0),\nonumber\\
&&{\mathbb{P}}(0)=\sum^{r_0}_{j=1}u_ju_j^T.\ear

A particular set of such eigenvectors $u_j$
can be constructed as follows.
In the subspace corresponding to $q=t_3'=0$ (if this occurs
at all) we take an arbitrary set of normalised real
eigenvectors $u_j$. For $q=0,\ t_3'\not=0$ we consider
the common eigenvectors of $Q'$ and $T_3'$ in
${\mathbb{C}}_{n-3}$
\bear\label{3.22a}
Q'u(0,t_3')&=&0,\nonumber\\
T_3'u(0,t_3')&=&t_3'u(0,t_3').\ear
{}From (\ref{3.13}) we get
\bear\label{3.22b}
Q'u^*(0,t_3')&=&0,\nonumber\\
T_3'u^*(0,t_3')&=&-t_3'u^*(0,t_3').\ear
This shows that $u^*(0,t_3')$ are eigenvectors to
$Q',T_3'$ with eigenvalues $(0,-t_3')$.
Therefore the set of vectors
\[u(0,t_3'),\quad u^*(0,t_3'),\quad (t_3'>0),\]
where we can choose the normalisation such that
\be\label{3.22c}
u^\dagger(0,t_3')u(0,t_3'')=\delta_{t_3',t_3''},\ee
forms a basis of eigenvectors for $q=0, t_3'\not=0$.
{}From (\ref{3.22b}) we see that we also have
\be\label{3.22d}
u^{*\dagger}(0,t_3')u(0,t_3'')=0.\ee
The vectors
\bear\label{3.22e}
u_1(0,t_3'):&=&\sqrt2\ {\RRe}\ u(0,t_3'),\nonumber\\
u_2(0,t_3'):&=&\sqrt2\ {\RIm}\ u(0,t_3'),\nonumber\\
\qquad &&(t_3'>0),\ear
are then linearly independent, normalised vectors in
${\mathbb{R}}_{n-3}$ which we can choose as basis vectors satisfying
(\ref{3.22}). In this basis the real, antisymmetric matrix $\tilde T_3'$
has the following structure in the $q=0$ subspace:
It is block-diagonal, with possibly a number of zeros and
then $2\times 2$ matrices
\[\left(\begin{array}{cc}
0&t_3'\\
-t_3'& 0\end{array}\right)\]
where  $t_3'$ are the positive eigenvalues of the hermitian
matrix $T_3'$:
\be\label{3.22f}
{\tilde T}_3^{'(0)}=\left(\begin{array}{c|c|c}
\begin{array}{ccccc}
0&&&&\\
&.&&&\\
&&.&&\\
&&&.&\\
&&&&0\\
\end{array}
&0&0\\
\hline
0&\begin{array}{cc}
&\\ 0&t_3'\\
&\\
-t_3'&0\\
&\\
\end{array}& 0\\
\hline
0&0&\begin{array}{ccc}
.&&\\
&.&\\
&&.\end{array}\end{array}\right).\ee
We can consider (\ref{3.22f}) as a standard form for
$\tilde T_3^{'(0)}$, the submatrix of $\tilde T_3'$
in the $q=0$ subspace.

For $q\not=0$ we consider simultaneously
the subspaces $S_q$ and $S_{-q}$.
We can then without loss of generality assume $q>0$. Let
$u(q,t_3')$ be the common eigenvectors of $Q'$ and
$T_3'$ (cf.(\ref{3.11})) in ${\mathbb{C}}_{n-3}$
\bear\label{3.23}
Q'u(q,t_3')&=&qu(q,t_3'),\nonumber\\
T_3'u(q,t_3')&=&t_3'u(q,t_3'),\ear
where $t_3'$ runs over all eigenvalues of $T_3'$
corresponding to charge
$q$ and we have suppressed a possible degeneracy index.
{}From (\ref{3.13}) we get
\bear\label{3.31}
Q'u^*(q,t_3')&=&-qu^*(q,t_3'),\nonumber\\
T_3'u^*(q,t_3')&=&-t_3'u^*(q,t_3').\ear
With suitable numbering and phases, the vectors
$u^*(q,t_3')$ can thus be considered as the eigenvectors
of $Q',T_3'$ in $S_{-q}$ associated with the eigenvalue pair
$(-q,-t_3')$. We can normalise $u(q,t_3')$ to
\be\label{3.32}
u^\dagger(q,t_3')u(q,t_3'')=\delta_{t_3't_3''}\ee
and (\ref{3.31}) implies
\be\label{3.33}
u^{*\dagger}(q,t_3')u(q,t_3'')=0.\ee
We can define real vectors $u_{1,2}(q,t_3')$ by
\bear\label{3.34}
u_1(q,t_3'):&=&\sqrt2\cdot {\RRe}\ u(q,t_3'),\nonumber\\
u_2(q,t_3'):&=&\sqrt2\cdot {\RIm}\ u(q,t_3').\ear
{}From (\ref{3.32}), (\ref{3.33}) we find that
$u_{1,2}(q,t_3')$ are linearly independent real vectors in
${\mathbb{R}}_{n-3}$ satisfying
\bear\label{3.35}
u_1^T(q,t_3')u_1(q,t_3'')&=&\delta_{t_3't_3''},
\nonumber\\
u_2^T(q,t_3')u_2(q,t_3'')&=&\delta_{t_3't_3''},
\nonumber\\
u_1^T(q,t_3')u_2(q,t_3'')&=&0.
\ear

Let us now choose the basis vectors in ${\mathbb{R}}_{n-3}$ as follows.
In the  subspace $S_0$  we take the $u_j$ of (\ref{3.22}). In the
subspaces $S_q+S_{-q}\ (q>0)$
we choose the vectors $u_{1,2}$ of (\ref{3.34}) and
denote them collectively by
\[u_{1,\alpha},\ u_{2,\alpha},\quad (\alpha=1,2,...r),\]
where $\alpha$ stands for the pair $(q,t_3')$ plus a possible
degeneracy index. We can then decompose any given vector
$\chi'\in {\mathbb{R}}_{n-3}$ as
\be\label{3.36}
\chi'=\sum^{r_0}_{j=1}\chi_j'u_j
+\sum^r_{\alpha=1}(\chi_{1,\alpha}'u_{1,\alpha}+
\chi_{2,\alpha}'u_{2,\alpha})\ee
and we get
\be\label{3.37}
\chi^{'T}\chi'=\sum^{r_0}_{j=1}(\chi_j')^2+\sum^r_{\alpha=1}
[(\chi_{1,\alpha}')^2+(\chi_{2,\alpha}')^2].\ee
The fields $\chi_{1,\alpha},\chi_{2\alpha}\quad(\alpha=1,...,r)$
corresponding to $q\not=0$ can  be rearranged into $r$ complex
Higgs fields. For this we define
\bear\label{3.38}
&&u_\alpha:=\frac{1}{\sqrt 2}(u_{1,\alpha}+iu_{2,\alpha}),\nonumber\\
&&\phi_\alpha':=\frac{1}{\sqrt 2}(\chi_{1,\alpha}'-i\chi_{2,\alpha}'
).\ear
The $u_\alpha$ are the complex
eigenvectors $u(q,t_3')$ of (\ref{3.23}) which we write now
as
\bear\label{3.39}
&&\tilde Q'u_\alpha=iq_\alpha u_\alpha,\nonumber\\
&&\tilde T_3'u_\alpha=it_{3,\alpha}' u_\alpha,\nonumber\\
&&(q_\alpha>0).\ear
We get then
\be\label{3.40}
\chi'=\sum^{r_0}_{j=1}\chi_j'u_j+\sum^r_{\alpha=1}(\phi_\alpha'u_\alpha
+\phi_\alpha^{'*}u_\alpha^*)\ee
\be\label{3.41}
\frac{1}{2}\chi^{'T}\chi'=\frac{1}{2}\sum^{r_0}_{j=1}\chi_j'\chi_j'
+\sum^r_{\alpha=1}\phi_\alpha^{'*}\phi_\alpha',\ee
\bear\label{3.42}
\frac{1}{2}\partial_\mu\chi^{'T}\partial^\mu\chi'&=&\frac
{1}{2}\sum^{r_0}_{j=1}\partial_\mu\chi_j'\partial^\mu\chi_j'
+\sum^r_{\alpha=1}\partial_\mu\phi_\alpha^{'*}\partial^\mu
\phi_\alpha',\ear
\bear\label{3.43}
{\cal L}^{(4)}_\chi&=&-e\Bigl\{A_\mu
J^\mu_{H,em}+\frac{1}{sc}Z_\mu J^\mu_{H,NC}\nonumber\\
&&+\frac{1}{\sqrt2s}\left(W^+_\mu J^\mu_{H,CC}+W^-_\mu J^{\mu\dagger}_{HCC}
\right)\Bigr\}\ear
where
\be\label{3.44}
W^\pm_\mu=\frac{1}{\sqrt2}(W^1_\mu\mp iW^2_\mu)\ee
and $J^\mu_{H,em}, J^\mu_{H,NC},J^\mu_{H,CC}$ are
the electromagnetic, neutral and charged Higgs currents,
respectively:
\be\label{3.45}
J^\mu_{H,em}=i\sum_\alpha q_\alpha\phi_\alpha^{'*}
\stackrel{\leftrightarrow}{\partial^\mu}\phi_\alpha',\ee
\be\label{3.46}
J^\mu_{H,NC}=\frac{1}{2}\sum_{j,k}\chi_j'\tilde T^{'(0)}
_{3,jk}\stackrel{\leftrightarrow}{\partial^\mu}\chi_k'
+i\sum_\alpha(t_{3,\alpha}'-s^2q_\alpha)\phi_\alpha^{'*}
\stackrel{\leftrightarrow}{\partial^\mu}\phi_\alpha',\ee
\bear\label{3.47}
J^\mu_{H,CC}&=&\frac{1}{2}\chi^{'T}(\tilde T_1'+i\tilde T_2')
\stackrel{\leftrightarrow}{\partial^\mu}
\chi',\nonumber\\
&&\nonumber\\
&&(\stackrel{\leftrightarrow}{\partial^\mu}=\stackrel{\rightharpoonup}
{\partial^\mu}-\stackrel{\leftharpoonup}{\partial^\mu})\ear.

The fields $\chi_j'$ ($j=1,..,r_0$) correspond
to neutral particles, and the fields $\phi_\alpha'$ annihilate
(create) particles of charge $eq_\alpha (-eq_\alpha)$.
In (\ref{3.45})-(\ref{3.47}) we have thus used a basis
where the electromagnetic current and the contribution of the charged
fields to the neutral current are diagonal. The part of the neutral current
from neutral fields is written in the basis where $\tilde T_3^{'(0)}$
is in the standard form (\ref{3.22f}).

Of course, the above basis is, in general, not identical to the
mass eigenbasis for the Higgs fields. The Higgs boson mass matrix ${\cal M}$
is obtained from the bilinear term in the potential $V$ (\ref{2.22})
expressed in terms of the shifted fields $\chi'$ (\ref{2.30}):
\be\label{3.48}
V(v+\chi')=V(v)+\frac{1}{2}\chi^{'T}{\cal M}{\chi'}+... .\ee

Let \[\hat\chi_j\qquad (j=1,...,r_0)\]
be the real fields diagonalising this mass matrix in the
neutral $(q=0)$ sector and let
\[\hat\phi_{q,\alpha}\qquad (\alpha=1,...,r_q)\]
be the complex mass eigenfields carrying charge $q>0$. We have then
\be\label{3.49}
\hat\chi\equiv\left(\begin{array}{c}
\hat\chi_1\\ \cdot\\ \cdot\\ \hat\chi_{r_0}\end{array}\right)=V^{(0)T}
\left(\begin{array}{c}
\chi_1'\\ \cdot\\ \cdot\\ {\chi'}_{r_0}\end{array}\right),\ee
\be\label{3.50}
\hat\Phi_q\equiv\left(\begin{array}{c}
\hat\phi_{q,1}\\ \cdot\\ \cdot\\ \hat\phi_{q,
r_q}\end{array}\right)=V^{(q)\dagger}
\left(\begin{array}{c}
\phi_{q,1}'\\ \cdot\\ \cdot\\ \phi_{q,r_q}'\end{array}\right),\ee
where $V^{(0)}$ is a real orthogonal $r_0\times r_0$
and the $V^{(q)}$ are unitary $r_q\times r_q$ matrices. Of course,
only Higgs fields of the same charge mix. Inserting (\ref{3.49}) and
(\ref{3.50}) in (\ref{3.45}), (\ref{3.46}), we get the following
expressions for the electromagnetic and neutral Higgs currents
in terms of mass eigenfields:
\be\label{3.51}
J^\mu_{H,em}=i\sum_{q>0}q\hat\Phi_q^\dagger\stackrel{\leftrightarrow}
{\partial^\mu}\hat\Phi_q,\ee
\be\label{3.52}
J^\mu_{H,NC}=\frac{1}{2}\sum_{j,k}\hat\chi_j\hat T_{3,jk}^{(0)}
\stackrel{\leftrightarrow}{\partial^\mu}\hat\chi_k
+i\sum_{q>0}\hat\Phi_q^\dagger(\hat T_3^{(q)}-s^2q)
\stackrel{\leftrightarrow}{\partial^\mu}\hat\Phi_q.\ee
Here we have for $q=0$
\be\label{3.53}
\hat T_3^{(0)}=V^{(0)T}\tilde T_3^{'(0)}V^{(0)},\ee
and for $q>0$
\be\label{3.54}
\hat T_3^{(q)}=V^{(q)\dagger}\cdot
{\rm diag}\ (t^{'(q)}_{3,1},...,t^{'(q)}_{3,r_q})\cdot
V^{(q)}\ee
with $t^{'(q)}_{3,1}...,t^{'(q)}_{3,r_q}$ the eigenvalues
of $T_3'$ occurring for charge $q$.
The matrices $V^{(q)}\\
(q>0)$ are the analogues of the Cabibbo-Kobayashi-Maskawa (CKM)
matrix \cite{13} governing the quark-W-boson couplings.
There is still some freedom in the choice of
$V^{(q)}$, i.e. we can make the replacement
\be\label{3.55}
V^{(q)}\longrightarrow U_1^{(q)\dagger}V^{(q)}U_2^{(q)},\ee
where $U_1^{(q)}$ is a unitary matrix commuting with
${\rm diag}\ (t_{3,1}^{'(q)},...,t^{'(q)}_{3,r_q})$ and $U_2^{(q)}$ is
a unitary matrix commuting with the mass matrix for charge $q$.
In particular if, and only if, all mass eigenvalues
and all $t^{'(q)}_{3,\alpha}$
corresponding to charge $q$ are different then $U_{1,2}^{(q)}$
are  diagonal unitary matrices (apart from
trivial renumbering of the fields). Thus the questions concerning diagonal
versus non-diagonal real/complex couplings in the neutral current
involving physical Higgs fields can immediately be answered:
We find the following for charged physical Higgs fields:

(1) A non-diagonal $Z$-Higgs-Higgs boson coupling requires at least
two Higgs fields with the same charge, but different mass,
being linearly related to fields with two different eigenvalues
$t^{'(q)}_{3,\alpha}$ of the matrix $T_3'$ in the sector
corresponding to charge $q$. Here $T_3'$ is related by a
projection (cf. (3.8) ff.) to the matrix of the
third component of the weak isospin.

(2) If only two charged  fields mix, the mixing
matrix $V^{(q)}$ can always be made real by a replacement
(\ref{3.55}). Thus, in this  this case, the non-diagonal $Z$-Higgs-Higgs
boson coupling can, without loss of generality, be assumed to be real.

(3) A non-diagonal complex $Z$-Higgs-Higgs coupling (whose phase(s)
cannot be rotated away) for charged Higgs bosons requires at least
three Higgs fields of the same charge with different masses, where
these fields are linearly related to fields with at least three
different eigenvalues $t^{'(q)}_{3,\alpha}$. For 3 Higgs fields
the form
of the mixing matrix $V^{(q)}$ can be chosen
in analogy to  the CKM matrix for 3 quark generations.

In the neutral Higgs sector the $Z\hat\chi\hat\chi$ coupling
matrix $\hat T_3^{(0)}$ (\ref{3.53})
is, in general, an arbitrary
antisymmetric matrix. Indeed, $V^{(0)}$
in (\ref{3.49}) can be an arbitrary orthogonal matrix and any
antisymmetric matrix $\hat T_3^{(0)}$ can be brought to the
standard form (\ref{3.22f}) by an orthogonal
transformation. On general grounds we did not find a restriction
on the possible values of $t_3'$.

The charged scalar current (\ref{3.47}) can also be straightforwardly
expressed in terms of the physical Higgs fields. From
(\ref{3.10a}) we have
\be\label{3.56}
[\tilde Q',\tilde T_1'\pm i\tilde T_2']=\pm i(
\tilde T_1'\pm i\tilde T_2').\ee
This guarantees that only fields differing by one unit of charge
couple in $J^\mu_{H,CC}$, as it must be by charge conservation.
Otherwise we did not find any useful general statement for this
current.

\section{A model with one Higgs singlet and an arbitrary number
of Higgs doublets}
\setcounter{equation}{0}

In this section we consider as a specific  example a model with
one complex Higgs singlet $\phi_0$ of hypercharge
$y=1$ and $l$ doublets $\phi_j(j=1,...,l)$ of
hypercharge $y=1/2$. The Higgs boson Lagrangian (\ref{2.22})
is then
\be\label{4.1}
{\cL}_\chi=(D_\mu\phi_0)^\dagger D^\mu\phi_0
+\sum^l_{j=1}(D_\mu\phi_j)^\dagger(D^\mu\phi_j)-V,\ee
\bear\label{4.1a}
D_\mu\phi_0&=&(\partial_\mu+ig'B_\mu)\phi_0,\nonumber\\
D_\mu\phi_j&=&(\partial_\mu+ig W^a_\mu\frac{1}{2}\tau^a
+ig'\frac{1}{2}B_\mu)\phi_j,\nonumber\\
&&(j=1,...,l),\ear
\be\label{4.1b}
\phi_j={\phi_{1/2,j} \choose \phi_{-1/2,j}},\ee
\be\label{4.2}
V=V_2+V_3+V_4,\ee
\bear\label{4.3}
V_2&=&\mu\phi^\dagger_0\phi_0
+\sum^l_{j,k=1}\lambda_{jk}\phi^\dagger_j\phi_k,\nonumber\\
&&(\mu^*=\mu,\quad \lambda^*_{jk}=\lambda_{kj}),\ear
\bear\label{4.4}
V_3&=&\sum^l_{j,k=1}(\kappa_{jk}\phi^\dagger_0\phi_j^T\epsilon
\phi_k\ +\ {\rm h.c.}),\nonumber\\
&&(\kappa_{jk}=-\kappa_{kj},\quad \epsilon\ {\rm as\ in}
\ (2.10)),\ear

\bear\label{4.5}
V_4&=&\eta_0(\phi^\dagger_0\phi_0)^2+\sum^l_{j,k=1}\eta_{jk}\phi^\dagger
_j\phi_k\phi_0^\dagger\phi_0\nonumber\\
&&+\sum^l_{j,k,r,s=1}\left[\xi_{jkrs}
(\phi^\dagger_j\epsilon\phi^{\dagger T}_k)
(\phi^T_r\epsilon\phi_s)+\zeta_{jkrs}(\phi^\dagger_j\tau^a
\epsilon\phi_k^{\dagger T})(\phi^T_r\epsilon\tau^a\phi_s)\right]
\nonumber\\
&&(\eta^*_0=\eta_0;\ \eta^*_{jk}=\eta_{kj};\nonumber\\
&&\xi_{jk,rs}=-\xi_{kj,rs}=-\xi_{jk,sr}=\xi^*_{sr,kj};\nonumber\\
&&\zeta_{jk,rs}=\zeta_{kj,rs}=\zeta_{jk,sr}=\zeta^*_{sr,kj}).\ear
We note that the general form of the Lagrangian (\ref{4.1}) is
unchanged if we make a $U(l)$ transformation on the $l$ Higgs
doublets:
\be\label{4.6}
\phi_j \quad \longrightarrow \quad U_{jk} \phi_k \, .
\ee

The vacuum expectation values of
the Higgs fields are

\bear\label{4.8}
&&<0|\phi_0|0>=0,\nonumber\\
&&<0|\phi_j|0>={0 \choose v_j},
\nonumber\\
&&(j=1,...,l),\ear
where the $v_j$ are  complex numbers in general.
But by a $U(l)$ transformation (\ref{4.6}) we can always rotate
the Higgs fields such that only $v_1$ differs from zero and $v_1>0$. Then
$v_1$ must have the SM value
$\sqrt2v_1=(\sqrt2 G_F)^{-1/2}=246$ GeV
where $G_F$ is Fermi's constant.

The gauge condition (\ref{2.37}) reduces then to a condition
for  the first Higgs doublet $\phi_1$ -- as in the SM.  Thus
we get the following set of Higgs fields after spontaneous symmetry
breaking: $l$ complex fields of charge 1:
\be\label{4.10}
\Phi=\left(\begin{array}{c}
\phi_0\\ \phi_{1/2,2}\\ .\\ .\\ .\\ \phi_{1/2,l}\end{array}\right),
\ee
and $2l+1$ real fields of charge $0$:
\be\label{4.11}
\chi'=\left(\begin{array}{c}
\sqrt2 {\RRe} (\phi_{-1/2,1}-v_1)\\
\sqrt2 {\RRe}\phi_{-1/2,2}\\
\sqrt 2{\RIm}\phi_{-1/2,2}\\
.\\
.\\
\sqrt2 {\RRe}\phi_{-1/2,l}\\
\sqrt 2{\RIm}\phi_{-1/2,l}\end{array}
\right).\ee

In the following we will discuss mainly
the charged fields (\ref{4.10}) further. As we can easily see
from (\ref{4.2})-(\ref{4.5}) their mass matrix ${\cal M}^{(1)}$ has
the following structure:
\be\label{4.12}
{\cal M}^{(1)}=\left(\begin{array}{c|c}
{\cal M}^{(1)}_{11}&{\cal M}^{(1)}_{1k}\\
\hline
{\cal M}^{(1)}_{j1}& {\cal M}^{(1)}_{jk}\end{array}\right),
\qquad(2\leq j,k\leq l),\ee
where
\be\label{4.13}
{\cal M}^{(1)}_{1k}=-2v_1\kappa_{1k},\qquad(k=2,...,l).\ee
Thus, the $V_3$ term of the potential in (\ref{4.4})
induces a mixing of the singlet with the charged components
of the doublet Higgs fields.

Let us now consider the neutral current $J^\mu_{H,NC}$ (\ref{3.46})
in our model. In the basis (\ref{4.10}), (\ref{4.11})
it reads
\be\label{4.14}
J^\mu_{H,NC}=\frac{1}{2}\chi^{'T}\tilde T_3^{'(0)}
\stackrel{\leftrightarrow}{\partial^\mu}\chi'
+i\Phi^\dagger(T_3^{'(1)}-s^2)\stackrel{\leftrightarrow}{\partial^\mu}
\Phi,\ee
where
\bear\label{4.15}
\tilde T_3^{'(0)}&=&\left(\begin{array}{c|c|c|c}
0& 0& 0& 0\\
\hline
0&\begin{array}{cc} 0& +1/2\\ -1/2& 0\end{array}&0&0\\
\hline
0& 0& \begin{array}{ccc}
.&&\\ &.&\\ &&.\end{array}& 0\\
\hline
0& 0& 0& \begin{array}{cc} 0& +1/2\\
-1/2& 0\end{array}\end{array}\right),\nonumber\\
&&\nonumber\\
T_3^{'(1)}&=&\left(\begin{array}{c|c}
0&0\\
\hline
0&\frac{1}{2}\delta_{jk}\end{array}\right),\qquad
(2\leq j,\ k\leq l).\ear
Following the discussion in section 3, we transform now to the
mass eigenbasis of the Higgs fields according to (\ref{3.49}),
(\ref{3.50}):
\bear\label{4.16}
\hat\chi&=&V^{(0)T}\chi',\nonumber\\
\hat\Phi&=&V^{(1)\dagger}\Phi.\ear
Since $T^{'(1)}$ in (\ref{4.15}) has only 2 different eigenvalues,
the mixing problem for the charged fields
is as simple as in the case of 2 fields. It is easily shown
(cf. Appendix C) that $V^{(1)}$ can always be chosen to be a real
matrix. Then in terms of the mass eigenfields the neutral current
reads
\bear\label{4.17}
J^\mu_{H,NC}&=&\frac{1}{2}\hat\chi^T\hat T_3^{(0)}\stackrel
{\leftrightarrow}
{\partial^\mu} \hat\chi+i\hat\Phi^\dagger(\hat T_3^{(1)}-s^2)
\stackrel{\leftrightarrow}
{\partial^\mu} \hat\Phi,\nonumber\\
\hat T_3^{(0)}&=&V^{(0)T}T_3^{'(0)}V^{(0)},\nonumber\\
\hat T_3^{(1)}&=&V^{(1)\dagger}T_3^{'(1)}V^{(1)},\nonumber\\
&=&\left(\frac{1}{2}\delta_{jk}-\frac{1}{2}V^{(1)}_{1j}V^{(1)}_{1k}
\right),\nonumber\\
&& (1\leq j,\ k\leq l).\ear
In this model we get thus in accordance with our general
discussion non-diagonal real $Z$-Higgs-Higgs boson couplings for
the physical Higgs fields. In the next section we will use
this model in order to discuss the possibility of producing
at one-loop level a CP-violating coupling $Zb\bar bG$ which is
chirality-conserving and not suppressed by a factor proportional
to the $b$-mass.

\section{A chirality-conserving CP-violating $Zb\bar bG$ coupling}
\setcounter{equation}{0}

In \cite{14} the possibility of obtaining an effective CP-violating
and chirality-conserving coupling $Zb\bar bG$
\be\label{5.1}
{\cL}_{eff, CP}(x)=\bar b(x)T^a\gamma^\nu[h_{Vb}+h_{Ab}\gamma_5]
b(x)Z^\mu(x)G^a_{\mu\nu}(x)\ee
in renormalizable theories at one-loop level was discussed.
Here $b,Z^\mu$ are the $b$ quark and $Z$ boson fields, $T^a=\lambda^a/2$
are the generators of $SU(3)_c$ and $G^a_{\mu\nu}$ is the gluon field
strength tensor. In particular it was shown in \cite{14}
that in suitable models, called type I and II, the effective
couplings $h_{Vb}, h_{Ab}$ remained nonzero in the limit $m_b
\to0$. In this section we give an explicit example of a type I model
based on the discussion in sect. 4.

Let us start by writing down the most general
$SU(2)\times U(1)$-invariant Yukawa interaction for quarks
 in the model of Section 4:
\be\label{5.2}
{\cL}_{Yuk}=\sum^l_{j=1}\sum^3_{\alpha,\beta=1}\Bigl\{
-\bar d_{\alpha R}'C^j_{\alpha\beta}\phi^\dagger_j
{u_\beta \choose d_\beta'}_L+
\bar u_{\alpha R}C_{\alpha\beta}^{'j}\phi^T_j\epsilon
{u_\beta \choose d_\beta'}_L + {\rm h.c.}\Bigr\}\ee
Here $\alpha, \beta =1,2,3$ are generation indices,
$C^j_{\alpha\beta}$ and $C_{\alpha\beta}^{'j}$ are
arbitrary complex numbers, $u_\alpha,
d_\alpha'$
denote  $u$-type and $d$-type fields in the  weak
isospin basis, and $q_{R,L}= (1\pm\gamma_5)q/2$.
After spontaneous symmetry breaking and transformation to
the mass eigenbasis for the quark fields
(\ref{5.2}) reduces to
\bear\label{5.4}
{\cL}_{Yuk}&=&-\sum^3_{\alpha=1}\left[m_{d\alpha}\bar d_\alpha d_\alpha
+m_{u\alpha}\bar u_\alpha u_\alpha\right](1+\frac{\chi_1'}{
v_1\sqrt2})\nonumber\\
&& +\sum^l_{j=2}\sum^3_{\alpha,\beta=1}
\Bigl\{-\sum^3_{\rho=1}
\bar d_{\rho R} V^\dagger_{\rho\alpha}C^j_{\alpha\beta}
\phi^\dagger_j\left(\begin{array}{cc}u_\beta\\ \sum^3_{\gamma=1}
V_{\beta\gamma}
d_\gamma\end{array}\right)_L\nonumber\\
&&+\bar u_{\alpha R}
C^{'j}_{\alpha\beta}\phi^T_j\epsilon\left(\begin{array}{c}
u_\beta\\ \sum^3_{\gamma=1}V_{\beta\gamma}
d_\gamma\end{array}\right)_L
\ +\ {\rm h.c.}\Bigr\}.\ear
Here $V=(V_{\beta\gamma})$ is the CKM matrix and $C$, $C'$ denote the
$U(l)$-transformed Yukawa coupling matrices.

The general Yukawa interaction (\ref{5.4}) leads to flavour-changing
neutral currents (FCNC). In order to comply with experimental
bounds on FCNC processes one may either impose an appropriate
 discrete symmetry on ${\cL}_{Yuk}$ or fine-tuning of $C$, $C'$ is required.
Here our aim is to demonstrate a certain property of the Yukawa
couplings of charged Higgs bosons to the third quark-generation,
namely eq. (\ref{5.11}) below. For this purpose we discuss, as an example, 
 a model where only the right-handed top quark
couples to all the physical Higgs fields. This is realized
by setting
\bear\label{5.5}
C^j_{\alpha\beta}&=&0,\nonumber\\
C^{'j}_{\alpha\beta}&=&-\frac{m_t}{v_1}\delta_{\alpha3}\delta_{\beta3}
\beta_j',\nonumber\\
&&(j=2,...,l).\ear
where $m_t$ is the top quark mass and $\beta_j'$ are arbitrary complex
numbers. This leads to
\bear\label{5.6}
{\cL}_{Yuk}&=&-\sum^3_{\alpha=1}
\left[m_{d\alpha}\bar d_\alpha d_\alpha+m_{u\alpha}\bar u_\alpha u_\alpha
\right](1+\frac{\chi_1'}{v_1\sqrt2})\nonumber\\
&&-\sum^l_{j=2}\left\{\frac{m_t}{v_1}\beta_j'\bar t_R
\left[\phi_{1/2,j}\sum^3_{\alpha=1}V_{3\alpha}d_{\alpha L}-
\phi_{-1/2,j}t_L\right]\ +\ {\rm h.c.}\right\}\ear
In this way we have no flavour-changing neutral interactions
at tree level. Transforming to the mass eigenbasis for the Higgs
fields according to (\ref{4.16}) gives:
\bear\label{5.7}
{\cL}_{Yuk}&=&-\sum^3_{\alpha=1}\left[m_{d\alpha}
\bar d_\alpha d_\alpha+m_{u\alpha}\bar u_\alpha u_\alpha\right]
\left[1+\frac{1}{v_1\sqrt2}\sum^{2l-1}_{r=1}V^{(0)}
_{1r}\hat\chi_r\right]\nonumber\\
&&-\sum^l_{i=1}\left\{\frac{m_t}{v_1}\hat\beta_i\bar t_R
\hat\Phi_i\sum^3_{\alpha=1}V_{3\alpha}d_{\alpha L}\ +\ {\rm h.c.}
\right\}
\nonumber\\
&&+\sum^{2l-1}_{r=1}\left\{\frac{m_t}{v_1}\tilde\beta_r
\bar t_R\hat\chi_r t_L\ +\ {\rm h.c.}\right\};\ear
\be\label{5.8}
\hat\beta_i=\sum^l_{j=2}\beta_j'V^{(1)}_{ji},\quad(i=1,...,l);\ee
\bear\label{5.9}
\tilde\beta_r&=&\frac{1}{\sqrt2}\sum^l_{j=2}\beta_j'
\left(V^{(0)}_{2j-2,r}
+i V^{(0)}_{2j-1,r}\right),\nonumber\\
&&(r=1,...,2l-1).\ear

In order to compare with (7) of \cite{14} let us just look at the
$\phi tb$ coupling implied by (\ref{5.7}). We get
\bear\label{5.10}
{\cL}_{\phi tb}&=&-\frac{m_t}{v_1}\sum^l_{i=1}\beta_i\bar t_Rb_L\hat
\Phi_i\ + \ {\rm h.c.},\nonumber\\
\beta_i&=& \hat\beta_iV_{33}=\sum^l_{j=2}\beta_j'V_{33}V^{(1)}_{ji}.\ear
It was shown in \cite{14} that in models of the type
considered here one  gets nonzero effective CP-violating couplings
(\ref{5.1}) at the one-loop level, provided that
\be\label{5.11}
{\RIm} \beta_i\beta_j^*\not=0\quad {\rm for\ some}\ i\not=j,\ee
and the  corresponding Higgs masses are non-degenerate.
Clearly, in view of the large parameter space of the potential $V$
(\ref{4.2}), there is no reason why charged Higgs
bosons should be mass-degenerate.
Let us see if we can satisfy also (\ref{5.11}).
For  $l=2$ we get
two charged physical Higgs fields and (\ref{5.10}) gives
\bear\label{5.12}
\beta_1&=&\beta_2'V_{33}V^{(1)}_{21},\nonumber\\
\beta_2&=&\beta_2'V_{33}V^{(1)}_{22}.\ear
Because the $V_{ij}^{(1)}$ can be chosen to be real without loss of
generality (cf. Appendix C) we get
\be\label{5.12'}
{\RIm}\ \beta_1\beta_2^*=0 \ee
for arbitrary
complex $\beta_2', V_{33}$.
Thus, no CP-violating effective couplings (\ref{5.1}) can be
induced in this model.

For $l=3$, i.e., in a model with  three
charged physical Higgs fields the parameters $\beta_i$ are
\bear\label{5.13}
&&\beta_1=\beta_2'V_{33}V^{(1)}_{21}+\beta_3'V_{33}V_{31}^{(1)},\nonumber\\
&&\beta_2=\beta_2'V_{33}V^{(1)}_{22}+\beta_3'V_{33}V_{32}^{(1)},\nonumber\\
&&\beta_3=\beta_2'V_{33}V^{(1)}_{23}+\beta_3'V_{33}V_{33}^{(1)}.
\ear
Because $\beta_2'$ and $\beta_3'$ are arbitrary complex numbers  it
is now easy to realize (\ref{5.11}), e.g.,
\be\label{5.14}
{\RIm}\beta_1\beta_2^*\not=0.\ee

Thus in models where we start with one charged $SU(2)$ Higgs singlet
and at least three Higgs doublets we can in general
have effective CP-violating, chirality-conserving couplings of the
type (\ref{5.1}) which remain nonzero  for $m_b\to0$. For
the calculation of such couplings and a discussion of their
magnitudes  we refer to \cite{14}.

\section{Conclusions}

In this article we have analysed,
for gauge theories with gauge group $G = SU(2)\times U(1)$
being spontaneously broken to the electromagnetic $U(1)_{em}$  group,
some general properties
of the coupling of scalar fields to the electroweak gauge bosons
$W^\pm, Z$. We allowed the
scalar fields to carry arbitrary representations of $G$.
We found the following
general results.

The structure of the $Z$-scalar-scalar coupling is determined by the
charge matrix $Q'$ and the matrix $T_3'$ of the third component
of weak isospin, but both restricted to the space of physical
scalars. We discussed the scalar field basis for which
$T_3'$ is diagonal in the charge $q\not=0$ sectors and has
a standard form (\ref{3.22f}) in the $q=0$ sector. This basis is
in general not the mass eigenbasis and the rotation from the
former to the latter led us to the introduction
of orthogonal respectively unitary rotation matrices similar to the CKM matrix.
We found that non-diagonal (complex) $Z$-charged scalar couplings
in the mass eigenbasis require at least two (three) different
eigenvalues of $T_3'$ in the corresponding charge sector.

Finally we investigated models with one charged Higgs singlet
and any number $l$ of Higgs doublets. In these models the $Z$-charged
Higgs couplings in the mass eigenbasis can  always be made real
by a suitable rotation of fields. We considered then the coupling
of these fields to quarks and gave examples of models where no
flavour-changing neutral interactions at tree level occur.
We showed that
for $l\geq 3$ these models satisfy all requirements to
have CP-violating and chirality-conserving effective $Zb\bar bG$ couplings
(\ref{5.1}) at the one-loop level as investigated in detail in
\cite{14}. Such couplings are then not suppressed by factors
containing small quark masses. Thus further experimental search for such
CP-violating couplings, which have been considered theoretically
in \cite{14}-\cite{16} and experimentally in \cite{17,18},
should be quite interesting. Nonzero couplings of this kind
would point to a rich structure in the scalar sector as shown
in this article.

\section*{Acknowledgements}
The authors would like to thank A. Brandenburg and P. Haberl
for discussions, and B. Stech and Ch. Wetterich for discussions
and reading the manuscript. W. B. wishes to thank the Theory Division
at CERN for the hospitality extended to him.

\section*{Appendix A: Some properties of the $SU(2)\times U(1)$
representation carried by the Scalar Fields}

\renewcommand{\theequation}{A.\arabic{equation}}
\setcounter{equation}{0}
Consider the real representation (\ref{2.15})
of $G=SU(2)\times U(1)$ in the
space of the real $n$-component Higgs fields $\chi(x)
\in {\mathbb{R}}_n$ for each $x$ (cf. (\ref{2.14}).
We can trivially embed ${\mathbb{R}}_n$ in the complex $n$-dimensional
space ${\mathbb{C}}_n$ and consider the orthogonal representation (\ref{2.15})
of $G$ as unitary representation of
$G$ in ${\mathbb{C}}_n$:\\

\newpage
\bear\label{A.2}
&&R(U):{\mathbb{C}}_n\to {\mathbb{C}}_n,\nonumber\\
&&R^\dagger(U)R(U)={\eins}.\ear
For the representation $U\to R(U)$, considered as unitary
representation of $G$, all the standard results apply:
It can be reduced completely.
The hermitian generators are
\bear\label{A.3}
&&T_a=\frac{1}{i}\tilde T_a,\quad(a=1,2,3);\nonumber\\
&&Y=\frac{1}{i}\tilde Y.\ear
The irreducible parts of the representation are characterised by
$(t,y)$ where $t\in \{0,1/2,1,...\}$ with
$t(t+1)$ and $y$ the eigenvalues of
\be\label{A.4}
T_aT_a=-\tilde T_a\tilde T_a,\quad{\rm and}\quad
Y,\quad{\rm respectively}.\ee

Let $y_1,...,y_n$ be the eigenvalues of $Y$. Clearly, the eigenvalues
of
\be\label{A.5}
Y^2=-\tilde Y\tilde Y=\tilde Y^T\tilde Y\ee
are then $y^2_j\ (j=1,...,n)$, as introduced after (\ref{2.21}).

Now we want to discuss whether the representation $U\to R(U)$ in
(\ref{A.2}) is single- or multiple-valued. For the $SU(2)$ part of
the group $G$ there is no problem, since $SU(2)$ is singly
connected. But the $U(1)$ part of $G$ is multiply connected.
The representation matrices of the group elements $U(0,\psi)$ in
(\ref{2.1}) corresponding to the $U(1)$ factor of $G$ are:
\bear\label{A.6}
U(0,\psi)&=&e^{i\psi y_0}\to \exp(i\psi Y)\nonumber\\
&=& A^\dagger {\rm diag}\ (e^{i\psi y_1},...,e^{i\psi y_n})A.\ear
Here $A$ is the matrix which diagonalises $Y$ in ${\mathbb{C}}_n$.
With the condition (\ref{2.210}) we have
\bear\label{A.7}
&&e^{i\psi y_j}|_{\psi=\pi y^{-1}_0}=e^{i\psi y_j}|_{\psi
=-\pi y^{-1}_0}\nonumber\\
&&(j=1,...,n)\ear
and thus the representation is single-valued:
 There is a single
element $R(U)$ which corresponds to the element
$U(0,\pm\pi)=(-{\eins})\in G$.

We will now show that the representation $U\to R(U)$, considered
as a unitary representation in ${\mathbb{C}}_n$ has the following property:
If the irreducible representation $(t,y)$ appears
in the decomposition of the representation, then also $(t,-y)$ must
occur. Furthermore the irreducible representations $(t,y)$ and
$(t,-y)$ must have the same multiplicity. To prove this,
we note that due to (\ref{2.18}) and (\ref{A.3}) we have
\bear\label{A.8}
(T_aT_a)^T&=&(T_aT_a),\nonumber\\
Y^T&=&-Y.\ear
Consider now the double resolvent:
\be\label{A.9}
\frac{1}{(\xi-T_aT_a)(\eta-Y)}=
\sum_{t,y}\frac{{\mathbb{P}}(t,y)}{[\xi-t(t+1)](\eta-y)},\ee
where $\xi,\eta$ are arbitrary complex numbers and ${\mathbb{P}}(t,y)$ is
the projector onto the subspace of ${\mathbb{C}}_n$ carrying
the representation $(t,y)$ of $G$. (The irreducible representation
$(t,y)$ may occur with multiplicity one or higher).
We have
\bear\label{A.10}
\sum_{t,y}{\mathbb{P}}(t,y)&=&{\eins},\nonumber\\
{\mathbb{P}}(t,y)^\dagger&=&{\mathbb{P}}(t,y).\ear
{}From (\ref{A.8}) we get
\be\label{A.11}
\left[\frac{1}{(\xi-T_aT_a)(\eta-Y)}\right]^T=
\frac{1}{(\xi-T_aT_a)(\eta+Y)}\ee
which leads to
\be\label{A.12}
\sum_{t,y}\frac{{\mathbb{P}}(t,y)^T}{(\xi-t(t+1))(\eta-y)}=
\sum_{t,y}\frac{{\mathbb{P}}(t,y)}{(\xi-t(t+1))(\eta+y)}.\ee
Comparing the location of the poles in $(\xi,\eta)$
and the residues on the r.h.s. and l.h.s. of (\ref{A.12})
we find that with $(t(t+1),y)$ also $(t(t+1),-y)$ must
be the location of a pole and
\be\label{A.13}
{\mathbb{P}}(t,-y)={\mathbb{P}}(t,y)^T.\ee
This shows that the representations $(t,y)$ and $(t,-y)$ occur
with the same multiplicity, q.e.d.

Next we want to show the following theorem: If the vacuum expectation
value $v$ (\ref{2.24}) satisfies
\be\label{A.14}
{\mathbb{P}}(t,y)v\not=0,\ee
then $y$ must be one of the eigenvalues of $T_3$ in the
representation $(t,y)$:
\be\label{A.15}
y\in\{-t,-t+1,...,t\}.\ee
The proof is as follows: We have for the hermitian
electric charge generator $Q=\frac{1}{i}\tilde Q$:
\bear\label{A.16}
Q=T_3+Y&=&(T_3+Y)\cdot{\eins}\nonumber\\
&=&(T_3+Y)\sum_{t,y}{\mathbb{P}}(t,y)\nonumber\\
&=&\sum_{t,y}(T_3+y){\mathbb{P}}(t,y).\ear
{}From $Qv=0$ (cf. (\ref{2.25}))
and the fact that $T_3$ commutes with ${\mathbb{P}}(t,y)$, we get
\bear\label{A.17}
Qv&=&\sum_{t,y}(T_3+y){\mathbb{P}}(t,y)v\nonumber\\
&=&\sum_{t,y}{\mathbb{P}}(t,y)(T_3+y){\mathbb{P}}(t,y)v=0,\nonumber\\
\Longrightarrow&&\nonumber\\
&&{\mathbb{P}}(t,y)(T_3+y){\mathbb{P}}(t,y)v=0 ,\nonumber\\
\Longrightarrow&&\nonumber\\
&&(T_3+y){\mathbb{P}}(t,y)v=0.\ear
Thus, if ${\mathbb{P}}(t,y)v\not=0$, then $y$ is
one of the eigenvalues of $T_3$ in the representation
$(t,y)$ of $G$, q.e.d.

{}From (\ref{A.13}) we get for all $(t,y)$:
\bear\label{A.18}
v^T{\mathbb{P}}(t,y)v&=&v^T{\mathbb{P}}(t,-y)v\nonumber\\
&=&v^\dagger{\mathbb{P}}(t,y)v\geq0.\ear

\section*{Appendix B: Properties of gauge orbits of scalar fields}
\renewcommand{\theequation}{B.\arabic{equation}}
\setcounter{equation}{0}

In this appendix we derive some properties of the gauge orbits
of our general $n$ component real scalar field $\chi$ in
relation to the unitary gauge condition (\ref{2.38}). In the following
${\mathbb{R}}_{n-3}$, as defined in (\ref{2.39}),
is the subspace of the real vectors $\chi$ satisfying the gauge condition
(\ref{2.38}).

\underline{Theorem 1:} If $\chi\in {\mathbb{R}}_{n-3}$,
then also $\chi_1\in {\mathbb{R}}_{n-3}$ where
\be\label{B.1}
\chi_1=R(U)\chi\ee
with $U$ an arbitrary element of the electromagnetic subgroup
$U_{em}(1)\subset G$.

Proof: For an arbitrary element $U\in U_{em}(1)$ we have with
$\tilde Q$ as in (\ref{2.20})
\bear\label{B.2}
U&=&\exp[i\tilde\psi(\frac{1}{2}\tau_3+y_0)],\nonumber\\
\chi_1&=&R(U)\chi\nonumber\\
&=&\exp(\tilde\psi.\tilde Q)\chi.\ear
Then, using the commutation relations (\ref{2.21}), we find
immediately
\be\label{B.3}
v^T\tilde T_a\chi_1=v^T\tilde T_a\exp(\tilde\psi\tilde Q)\chi=0,\ee
if $v^T\tilde T_a\chi=0$ holds, i.e. if $\chi\in{\mathbb{R}}_{n-3}$.
But (\ref{B.3}) means that $\chi_1\in {\mathbb{R}}_{n-3}$, q.e.d.

\underline{Theorem 2:} There is no further subgroup $\bar G\subset G$,
$\bar G\not= U_{em}(1)$, which leaves ${\mathbb{R}}_{n-3}$ invariant.
In other words: If for all $\chi\in {\mathbb{R}}_{n-3}$ also
$R(U)\chi\in {\mathbb{R}}_{n-3}$, then $U\in U_{em}(1)$.

Proof (indirect): Assume, on the contrary, that $\bar G$ is a subgroup
of $G, \bar G\not= U_{em}(1)$, which leaves ${\mathbb{R}}_{n-3}$
invariant. Then $\bar G$ contains at least one one-parameter
subgroup $\bar U(1)\not=U_{em}(1)$. We must then have the following
for the elements $\bar U$ of $\bar U(1)$
\bear\label{B.4}
R(\bar U)&=&\exp(\tilde\psi\tilde{\bar Q}),\nonumber\\
&&(\bar U\in \bar U(1)),\ear
where $\tilde{\bar Q}$ is the matrix representing the generator of
$\bar U(1)$:
\be\label{B.5}
\tilde{\bar Q}=r_a\tilde T_a+r_4\tilde Q\ee
with $r_a(a=1,2,3)$ and $r_4$ real
numbers and
\be\label{B.6}
(r_1,r_2,r_3)\not=(0,0,0),\ee
since $\bar U(1)\not=U_{em}(1)$ by assumption.
Furthermore $\bar U(1)$ leaves ${\mathbb{R}}_{n-3}$
invariant, which means:
\be\label{B.7}
v^T\tilde T_aR(\bar U)\chi=v^T\tilde T_a\exp(\tilde
\psi\tilde{\bar Q})\chi=0\ee
for all $\chi\in {\mathbb{R}}_{n-3}$.

{}From (\ref{B.7}) we find by differentiating with respect
to $\tilde\psi$ at $\tilde \psi=0$:
\be\label{B.8}
v^T\tilde T_a\tilde{\bar Q}\chi=0\ee
for all $\chi\in {\mathbb{R}}_{n-3}$.

{}From the definition of ${\mathbb{R}}_{n-3}$ in (\ref{2.39}) we see
that (\ref{B.8}) can only hold if the vectors $v^T\tilde T_a\tilde
{\bar Q}$ are linearly dependent on $v^T\tilde T_b\ (b=1,2,3)$:
\be\label{B.9}
v^T\tilde T_a\tilde{\bar Q}=h_{ab}v^T\tilde T_b,\quad(h_{ab}\quad
{\rm real}).\ee
Multiplying by $v$ from the right and using (\ref{2.27})
we get
\be\label{B.10}
v^T\tilde T_a\tilde{\bar Q}v=0,\ee
\bear\label{B.11}
\Longrightarrow&&\nonumber\\
&&v^T\tilde T_a(r_b\tilde T_b+r_4\tilde Q)v=0.\ear
Since $\tilde Qv=0$ we get
\be\label{B.12}
v^T\tilde T_a(r_b\tilde T_bv)=0.\ee
Multiplying with $r_a\ (a=1,2,3)$ and summing yields:
\bear\label{B.13}
&&(v^T\tilde T_a^Tr_a)\cdot(r_b\tilde T_bv)=0,\nonumber\\
\Longrightarrow&&\ear
\be\label{B.14}
r_b\tilde T_bv=0.\ee

Since $\tilde T_bv$ are linearly independent (cf. (\ref{2.26})),
it follows that (\ref{B.14}) can only hold if $(r_1,r_2,r_3)=(0,0,0)$.
But this is a contradiction to (\ref{B.6}). Thus, the assumption
$\bar G\not= U_{em}(1)$ is disproved and theorem 2 holds.

Let us now define rest classes in $G$ with respect to $U_{em}(1)$:
\be\label{B.15}
U\sim U'\ee
if $U{U'}^{-1}\in U_{em}(1)$. Let $\hat G$ be the set formed by
these rest classes, i.e. the set of right cosets of $U_{em}(1)$.
A parametrization of $\hat G$ in a neighbourhood of the coset
of the unit element of $G$ is given by the elements of $SU(2)\subset
G$ (cf. (\ref{2.1}))
\be\label{B.16}
U(\vec\varphi,0)=\exp(i\frac{1}{2}\tau_a\varphi_a)\ee
with corresponding representation matrices
\be\label{B.17}
R(U(\vec\varphi,0))=\exp(\varphi_b\tilde T_b).\ee

In general the elements $U\in G$ which transform a given vector
$\chi\in {\mathbb{R}}_n$ into a vector $\chi_1\in {\mathbb{R}}_{n-3}$
\be\label{B.18}
R(U)\chi=\chi_1\in {\mathbb{R}}_{n-3}\ee
form isolated points in the coset space $\hat G$.

This can be shown as follows. Let $\chi$ be an arbitrary vector
from ${\mathbb{R}}_n$ and let $U_1\in G$ be a transformation
such that
\be\label{B.19}
R(U_1)\chi=\chi_1\in {\mathbb{R}}_{n-3}.\ee
A suitable parametrization for the cosets in a neighbourhood
of the coset of $U_1$ is given by the following elements of $G$:
\be\label{B.20}
U(\vec\varphi,0)\cdot U_1.\ee
We have to study the system of equations
\bear\label{B.21}
h_a(\vec\varphi):&=&v^T\tilde T_aR(U(\vec\varphi,0))R(U_1)\chi
\nonumber\\
&=&v^T\tilde T_aR(U(\vec\varphi,0)\chi_1\nonumber\\
&=&0\ear
near $\vec\varphi=0$. We have
\be\label{B.22}
\left.\frac{\partial}{\partial \varphi_b}h_a(\vec\varphi)\right|_{\vec\varphi=0}
=v^T\tilde T_a\tilde T_b\chi_1.\ee
The point $\vec\varphi=0$ is an isolated solution of (\ref{B.21})
if
\be\label{B.23}
\det(v^T\tilde T_a\tilde T_b\chi_1)\not=0.\ee
If this determinant equals zero, $\vec\varphi=0$ need not be
an isolated solution.

Let us define the set
\be\label{B.24}
M'=\{\chi_1|\chi_1\in{\mathbb{R}}_{n-3},\ \det (v^T\tilde T_a\tilde
T_b\chi_1)=0\}.\ee
The determinant being zero represents one algebraic equation for
the vectors $\chi_1\in {\mathbb{R}}_{n-3}$. Thus the dimension
of $M'$ can be at most $n-4$.

\underline{Theorem 3:} There exists a neighbourhood of the vacuum
expectation value $\chi_1=v$ in ${\mathbb{R}}_{n-3}$ which has no point in
common with $M'$.

Proof: Since $T_av\ (a=1,2,3)$ are linearly independent
(cf. (\ref{2.26})) we have
\be\label{B.25}
\det(v^T\tilde T_a\tilde T_bv)\not=0.\ee
By continuity we have then
\be\label{B.26}
\det(v^T\tilde T_a\tilde T_b\chi_1)\not=0\ee
for $\chi_1$ in a suitable neighbourhood of $v$ in ${\mathbb{R}}_{n-3}$,
q.e.d.

\underline{Theorem 4:} The manifold $M'$ is invariant
under the action of the electromagnetic group $U_{em}(1)$:
\be\label{B.27}
\chi_2=R(U)\chi_1\in M'\quad {\rm if}\quad \chi_1\in M'
\quad {\rm and}\quad U\in U_{em}(1).\ee

Proof: Indeed, for $U\in U_{em}(1)$ we have with the notation according
to (\ref{B.2}):
\bear\label{B.28}
R^T(U)\tilde T_aR(U)=&=&\exp (\tilde\psi\tilde Q^T)\tilde T_a\exp
(\tilde \psi\tilde Q)\nonumber\\
&=&\exp (-\tilde\psi\tilde T_3)\tilde T_a\exp
(\tilde \psi\tilde T_3)\nonumber\\
&=&D_{ab}(U')\tilde T_b,\ear
where $U'=\exp(i\tilde\psi \tau_3/2)$ and
$(D_{ab}(U'))$ is the matrix of the adjoint representation of $SU(2)$.
We get then:
\bear\label{B.29}
v^T\tilde T_a\tilde T_b\chi_2&=&v^TR(U)R^T(U)\tilde T_aR(U)R^T(U)
\tilde T_bR(U)\chi_1\nonumber\\
&=&D_{aa'}(U')D_{bb'}(U')v^T\tilde T_{a'}
\tilde T_{b'}\chi_1,\nonumber\\
\Longrightarrow&&\nonumber\\
&&\det(v^T\tilde T_a\tilde T_b\chi_2)=\det(v^T\tilde T_{a'}
\tilde T_{b'}\chi_1),
\ear
q.e.d.

Consider next the manifold $M\in{\mathbb{R}}_n$ of those Higgs fields
$\chi$ whose gauge orbits intersect ${\mathbb{R}}_{n-3}$
in $M'$:
\bear\label{B.30}
M&=&\{\chi|\chi\in{\mathbb{R}}_n, \ {\rm such\ that\ there\ exists}
\nonumber\\
&&U_1\in G\quad {\rm with}\quad R(U_1)\chi=\chi_1\in M'\}.\ear

We have then
\be\label{B.31}
\chi= R^{-1}(U_1)\chi_1,\quad \chi_1\in M'.\ee
Since $M'$ is invariant under the action of $U_{em}(1)$ it is sufficient to
choose for $U_1$ in (\ref{B.31}) only one representative of each right
coset of $U_{em}(1)$. This means that the parameters needed to describe
the manifold $M$ are those of $M'$ (at most $n-4$) and the
3 parameters (at most) of the coset space $\hat G$. Thus $M$ has
at most dimension $n-4+3=n-1$ and is a set of measure zero
in ${\mathbb{R}}_n$.

We summarize these findings as follows.

\underline{Theorem 5:} The elements $U\in G$ which transform
a given Higgs field $\chi\in {\mathbb{R}}_n$ (at a given space-time
point) into a vector $\chi_1$ satisfying the gauge condition (\ref{2.38})
belong for general $\chi$ to isolated points in the coset space
$\hat G$. The vectors $\chi\in {\mathbb{R}}_n$ where this is
not the case form a manifold $M$ of dimension $\leq n-1$ in
${\mathbb{R}}_n$ and thus a set of measure zero.

Finally, let us discuss the question of multiple intersections
of the gauge orbit of a vector $\chi$ with ${\mathbb{R}}_{n-3}$
(\ref{2.39}). For given $\chi\in{\mathbb{R}}_n$ we have
\[\chi_1=R(U_0)\chi\in{\mathbb{R}}_{n-3},\]
where $U_0\in G$ is the transformation that maximises the function
$f(U)$ (\ref{2.40}). It is clear that also the element
$U_0'\in G$ which minimises $f(U)$:
\[f(U)\geq f(U_0')\ {\rm for\ all}\ U\in G\]
leads to an intersection of the gauge orbit of a vector $\chi$ with
${\mathbb{R}}_{n-3}$:
\[\chi_2:=R(U_0')\chi\in{\mathbb{R}}_{n-3}.\]
The same holds true for all stationary points of $f(U)$.
Thus, in general, the gauge orbit of $\chi$ will have multiple
intersections with ${\mathbb{R}}_{n-3}$. Consequently the gauge
condition (\ref{2.38}) can and must be sharpened by restricting
$\chi$ to a region in ${\mathbb{R}}_{n-3}$ where the gauge orbits
have single intersections only. A suitable restriction is to the region
in ${\mathbb{R}}_{n-3}$ defined by the absolute maxima of the
functions $f(U)$:
\bear\label{B.32}
R_{n-3}'&=&\{\chi|\chi\in {\mathbb{R}}_{n-3},\nonumber\\
&&v^TR(U)\chi\leq v^T\chi\quad{\rm for\ all} \ U\in G\}
\ear
Here we assume that the absolute maxima of $v^TR(U)\chi$ have
no degeneracy (except for a set of measure zero).

In the SM with one Higgs doublet
the restriction of the form (\ref{B.32}) is, of course, well known.
Taking it into account in the
path integral quantisation by the method of Fadeev and Popov
\cite{19} one finds that even in the unitary
gauge  ghost fields are required. This was first demonstrated
in the canonical quantisation procedure by Weinberg \cite{7}.

\section*{Appendix C: Properties of the matrix $V^{(1)}$}
\renewcommand{\theequation}{C.\arabic{equation}}
\setcounter{equation}{0}

In this appendix we show that the matrix $V^{(1)}$ of
(\ref{4.16}) can always be chosen to be real. According to
(\ref{3.55}) we are free
to make the transformations
\be\label{C.1}
V^{(1)}\to U_1^\dagger V^{(1)}U_2\ee
where
\be\label{C.2}
U_2={\rm diag} (e^{i\psi_1},...,e^{i\psi_l}).\ee
and $U_1$ is a unitary $l\times l$ matrix commuting
with $T^{'(1)}_3$ (\ref{4.15}).
Thus $U_1$ must have the form
\be\label{C.3}
U_1=\left(\begin{array}{c|c}
e^{i\varphi_1}& 0\\
\hline
0&U_1'\end{array}\right)\ee
with $U_1'$ being a unitary $(l-1)\times (l-1)$ matrix.

With a suitable choice of $U_1$ and $U_2$ in (\ref{C.1})
we can achieve
\bear\label{C.4}
&&V^{(1)}_{11},V^{(1)}_{12},..., V^{(1)}_{1l}\ {\rm real},
\nonumber\\
&&V^{(1)}_{21}\ {\rm real},\ V^{(1)}_{31}=... = V^{(1)}_{l1}=0.\ear
Two cases can be distinguished:
(i) $V^{(1)}_{21}=0$.
Then applying (\ref{C.1}) we can immediately bring $V^{(1)}$
to the real form:
\be\label{C.5}
V^{(1)}=\left(\begin{array}{c|c}
V^{(1)}_{11}& 0\\
\hline
0&\eins_{l-1}\end{array}\right).\ee

(ii) $V^{(1)}_{21}\not=0$.
Then the unitarity relations for $V^{(1)}$ require
\[V^{(1)}_{22},...,V^{(1)}_{2l}\quad{\rm real}.\]
By a suitable choice of $U_1,U_2$ in (\ref{C.1})
we can then achieve
\be\label{C.6}
V^{(1)}_{32}\ {\rm real},\ V^{(1)}_{42}=...=V^{(1)}_{l2}=0\ee
and repeat the above reasoning (i) and (ii) with $V^{(1)}_{32}$
playing the role of $V^{(1)}_{21}$. In this way we see
that indeed $V^{(1)}$ can be made real by a transformation
(\ref{C.1}).

\newpage
\begin{center}{\bf Table 1}\\

\bigskip
\begin{tabular}{c|c|c|c|c}
$t$&$y$&$t(t+1)-y^2$&$y^2$&$[t(t+1)-y^2]/(2y^2)$\\
\hline
0&0&0&0&\\
1/2&$\pm1/2$&1/2&1/4&1\\
1&$\pm1$&1&1&1/2\\
1&0&2&0&$\infty$\\
3/2&$\pm3/2$&3/2&9/4&1/3\\
3/2&$\pm1/2$&7/2&1/4&7\\
2&$\pm2$& 2&4&1/4\\
2&$\pm1$& 5&1&5/2\\
2&0&6&0&$\infty$\\
5/2&$\pm5/2$&5/2&25/4&1/5\\
5/2&$\pm3/2$&13/2&9/4&13/9\\
5/2&$\pm1/2$&17/2&1/4&17\\
3&$\pm2$&8&4&1\\
3&$\pm1$&11&1&11/2\\
3&0&12&0&$\infty$\end{tabular}\\
\end{center}

\medskip
\noindent Values for the weak isospin $t$, the weak
hypercharge $y$ and the $\rho$ parameter (\ref{2.44}),
(\ref{2.45}) for the representations of $G=SU(2)\times U(1)$
with $t\leq 3$ which can give a nonzero contribution in the
sums (\ref{2.43}).

\end{document}